\documentclass[twocolumn,pre,showpacs,preprintnumbers,floatfix,amsmath,amssymb]{revtex4}

\usepackage{graphicx}

\begin{document}

\title{In phase and anti-phase synchronization of coupled homoclinic chaotic oscillators}

\author{I. Leyva$^{1,3}$}
\email{ileyva@ino.it}
\author{E. Allaria$^{1}$}
\author{S. Boccaletti$^{1}$}
\author{F.T. Arecchi$^{1,2}$}

\affiliation{$^1$ Istituto Nazionale di Ottica Applicata, Largo E. Fermi 6, 50125 Florence, Italy}
\affiliation{$^2$ Departement of Physics, University Of Firenze, Italy}
\affiliation{$^3$ Universidad Rey Juan Carlos. c/Tulipan  s/n. 28933 Mostoles Madrid, Spain}

\date{\today}

\begin{abstract}
We numerically investigate the dynamics of a closed chain of unidirectionally coupled
oscillators in a regime of homoclinic chaos. The emerging synchronization regimes show
analogies with the experimental behavior of a single chaotic laser subjected to a delayed feedback
\end{abstract}
 \maketitle

The study of collective phenomena in chains of chaotic
oscillators has recently attracted a wide interest, for the variety of possible
scenarios that can be found, and the analogies with biological complex systems.
Experiments have been performed on arrays of optical systems \cite{arraydiode1},
electronic circuits \cite{circuito}, neurons \cite{arrayneuro,arrayneuro2}
and chemical oscillators \cite{arraychem},
reporting different synchronization patterns, such as antiphase synchronization,
and clustering \cite{report}. In a theoretical work, the dynamics of an unidirectionally
coupled ring of chaotic systems has been explored
in view of possible applications to neural systems \cite{sanchez98}.
However, the need for parameter accurate control
has up to now limited experiments to a small number of oscillators, hence
most works on large chains are only numerical
.
In addition, an analogy between spatially extended systems
and delayed systems has already been drawn  \cite{STR}.

The aim of this paper is to show the equivalence
between the dynamics of one delayed system
and a unidirectionally coupled ring of identical systems.
A suitable framework for this symmetry makes it possible to retrieve
the dynamics of an arbitrary large array of coupled systems
by means of a single delayed system.

Our starting point is the recent report of the Delayed Self Synchronization (DSS) of  chaotic homoclinic spike trains
 in a CO$_2$ laser \cite{dss}. It has been shown that
this system displays a continuous return to a saddle focus where it
has a large sensitivity to external perturbations \cite{Allaria}.

\begin{figure}
\begin{center}
\includegraphics[width=8cm,height=9cm]{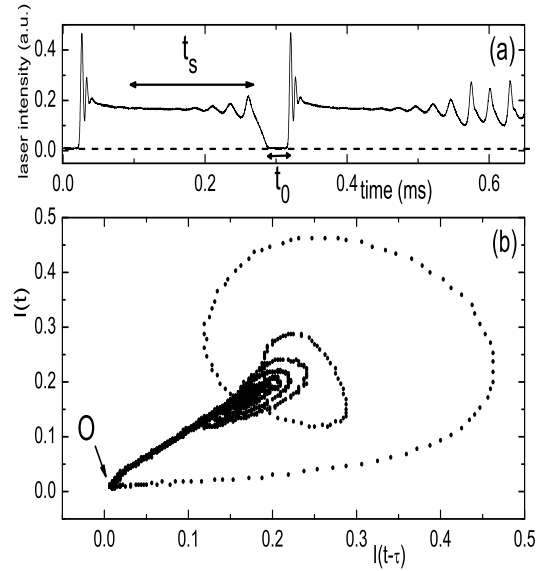}
\caption{(a) Experimental time intensity in the homoclinic regime. The dashed line corresponds to the baseline. (b) Phase space projection
 of the experimental time series.}
\label{homo}
\end{center}
\end{figure}

For convenience we report a stretch of the experimental time signal of  the homoclinic intensity (Fig. \ref{homo} (a)) as well as its
2D phase space projection. (Fig.  \ref{homo} (b)) obtained by an embedding technique and representing a superposition of a long
 sequence of spikes.  We realize that on each cycle the intensity returns to its zero value (baseline of Fig \ref{homo} (a) and
point O in Fig \ref{homo} (b)) emerging with a large spike followed by a decaying spiral toward a saddle region S. The escape from
S is represented by a growing oscillation, which appears as a nesty region in Fig \ref{homo} (b). Notice that the saddle region
S is very contracted in phase space but stretched in time-wise, viceversa the spike P occurs over a long time but is spread over a large 
space region.

The chaotic characteristic of the inter-spikes interval (ISI) is due to the chaotic permanence time t$_s$ around the saddle point S \cite{EPL},
whereas the permanence time t$_o$ on the baseline is fixed. This is confirmed by the values of the local Lyapunov exponents in the two regions
\cite{unpublished}.

In this condition,  a long-delayed feedback signal can stabilize
 complex periodic orbits of period $T$ characterized by a
pseudo-chaotic train of pulses, that is, by a limited sequence
with chaotic ISI, that repeats after a time T again for ever.

This ability of DSS can be of interest in relation to recent studies on neuronal transformation
mechanisms of short-term memories into permanent (long term) memories via the so
called synaptic reentry reinforcement \cite{srr}. Therefore, the possibility to describe
DSS as a synchronized state in a closed chain of oscillators is a relevant issue.

First it is important to establish under what conditions a chain of coupled oscillators can
be equivalent to a delayed system. Since the delay implies that the information propagates in one direction, just
unidirectional coupling will be considered in the oscillators chain, so that the oscillator at the site $i$ is driven
by the previous one at the site $i-1$. In addition,  the delayed reentry of the
signal in the system means that  the system is exposed to the total information
generated over a previous time stretch of size T$_d$. Meeting this condition imposes a closed boundary
with the last oscillator coupled to the first one.

With these boundary and coupling constraints, we build the array by using the scaled equations that model
the experimental laser system \cite{homo}
\begin{eqnarray}
\dot{x}_1^i&=& k_0x_1^i (x_2^i-1-k_1 \sin^2 x_6^i), \nonumber \\
\dot{x}_2^i&=& -\gamma_1 x_2^i- 2 k_0 x_1^i x_2^i+g x_3^i +x_4^i +p, \nonumber\\
\dot{x}_3^i&=& -\gamma_1 x_3^i+ g x_2^i +  x_5^i +p, \nonumber \\
\dot{x}_4^i&=& -\gamma_2 x_4^i+ z x_2^i +g x_5^i +z p,  \\
\dot{x}_5^i&=& -\gamma_2 x_5^i+ z x_3^i +g x_4^i +z p, \nonumber \\
\dot{x}_6^i&=& -\beta (x_6^i - b_0  + \frac{r  (x_1^i - \epsilon x_1^{i-1})}{1+\alpha x_1^i}) \nonumber,
\end{eqnarray}
where  the index $i$ denotes the $i^{th}$ site position, and for each oscillator
 $x_1$ represents the laser intensity, $x_2$ the
population inversion between the two resonant levels, $x_6$ the
feedback voltage which controls the cavity losses, while
$x_3,x_4$ and $x_5$ account for molecular exchanges between the
two levels resonant with the radiation field and the other
rotational levels of the same vibrational band.

In analogy with the experiment, the coupling on each oscillator
has been realized by adding a function of the intensity $(x_1)$  of the previous oscillator
to the equation of its feedback signal $x_6$ .

The parameters are
the same for all elements of the chain. Here, $k_0$ is the unperturbed cavity loss parameter,
$k_1$ determines the modulation strength,
$\gamma_1,\gamma_2, g$ are relaxation rates, $p_0$ is the
pump parameter, $z$ accounts for the number of rotational levels,
$\beta, b_0,r,\alpha$ are respectively the bandwidth, the bias voltage,
the amplification and the saturation factors of the feedback loop, and $\epsilon$ is the
coupling strength.
The values used in the numerical simulation to reproduce
the regime of homoclinic chaos \cite{homo}, are:
$k_0=28.5714$, $k_1=4.5556$, $\gamma_1=10.0643$,
$\gamma_2=1.0643$, $g=0.05$, $p_0=0.016$, $z=10$,
$\beta=0.4286$, $\alpha=32.8767$, $r=160$, $b_0=0.1032$.

The coupling strength $\epsilon$ is the control variable and  it can assume both negative
 and positive values, as in the experiment.  An important feature found in the experiment is that 
 the route towards the pseudo-chaotic state depends on the sign of the delayed feedback modulation.
If positive, the signal tends to reach in-phase synchronization with the delayed modulation,
while if it is negative, the coupling comes out to be phase-repulsive and the signal is in anti-phase with
the feedback perturbation.
\begin{figure}
\begin{center}
\includegraphics[width=8cm,height=9cm]{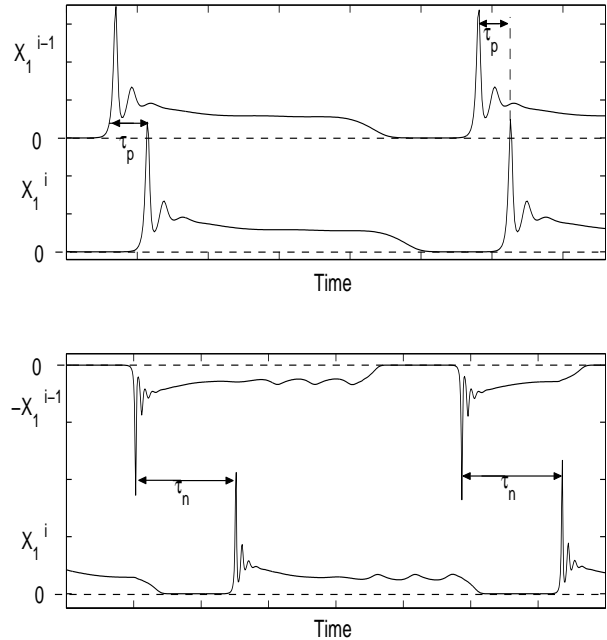}
\caption{Detail of the time profile of the $i^th$ (solid line) and $(i-1)^th$ elements of the chain, for
 (a) positive coupling and (b) negative coupling.}
\label{signdetail}
\end{center}
\end{figure}

This route is not symmetric for the coupling strength: a significantly stronger coupling is
needed to reach the pseudo-chaotic state for a positive than for a negative feedback
signal. This behavior is a consequence of the fact that the system
is more efficiently removed from the saddle focus neighborhood for a decrease
of the modulating signal. Eventually, when the coupling is strong enough, the system
becomes fully periodic for both positive and negative values,
i.e. there is no longer a pseudo-chaotic status.

The experimental data \cite{dss} reveal a small time offset between the modulation and
the signal, which is independent of the long delay $T_d$ and of  the coupling strength.
This offset depends on the sign of the modulation feedback,
and it has been measured to be $\tau_n$=140 $\mu$s for negative coupling and $\tau_p$=20 $\mu$s
for a positive, against an average ISI of 500 $\mu$s.

The numerical results explain how these asymmetries depend on the coupling sign.
In both cases, the slave intensity lags with
respect to the driving one, but by different amounts. Precisely, the driver's negative slope forces
 the slave to precipitate from the saddle region to the zero baseline. In case (a), the driver's and slave's
  collapses onto the zero baseline occur just one after the other and the negative oscillations around the spike
   do not induce a transition insofar as they occur while the slave is already on the baseline. On the contrary, in case (b)
  the first negative driver's slope which can force the slave to fall away from the saddle coincides with the first negative spike
  oscillation. As a result, considering that the permanence time on the baseline t$_o$ is a fixed fraction of the whole period,
  two different offsets $\tau_p < \tau_n$ occur.

Such an interpretation based on the numerical results can be also applied to the case of the DSS
where the delayed laser intensity stays for the driver signal.

Therefore we can say that the times $\tau_p$ and  $\tau_n$  correspond to different information propagation velocities
 along the  array,  with a lower velocity for the negative coupling.

\begin{figure}
\begin{center}
\includegraphics[width=8cm,height=6cm]{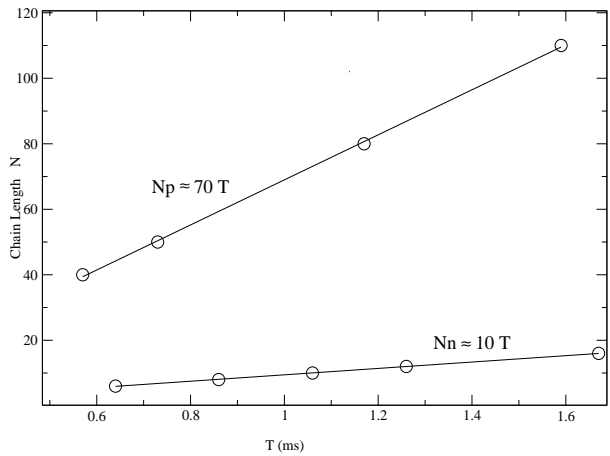}
\caption{Different response of the chain for negative (Nn, circles) and positive (Np, squares) coupling. The overall propagation
 velocities N$_j$/T (j=n,p) are in the ratio 1/7 as the two offsets time in Fig. \ref{signdetail}}
\label{linea}
\end{center}
\end{figure}

\begin{figure}
\begin{center}
\includegraphics[width=8cm,height=8cm]{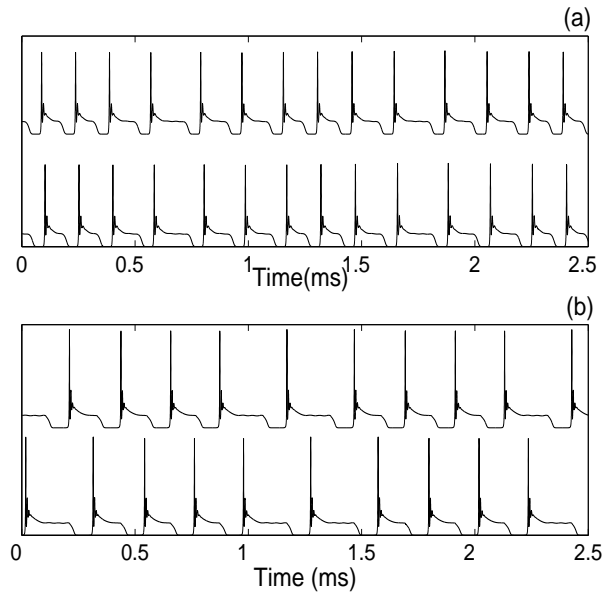}
\caption{Intensity profile of two neighbor oscillators  (a) $N=70$,
$\epsilon$=0.15, (b) $N=7$, $\epsilon$=-0.042. The data have been vertically shifted for a better display.}
\label{posneg}
\end{center}
\end{figure}

In order to model the laboratory system with an array of N oscillators, we must match the overall delay along the closed chain
N$\tau_j$ ( where j stays for $p$ or $n$ depending on the coupling) with a delayed time T$_d$ so that each system
 is exposed to a delayed version of its own signal  $x_1(t-T_d)=x_1(t-N\tau_j)$.
As  $\tau_p < \tau_n$, a longer chain will be need in a positive case to obtain the same periodicity.
The different propagation velocities can be seen in Fig.\ref{linea} where the relation
 between the repetition time $T$ and the oscillators number of a chain $N$ are reported for the two conditions.
From Fig. \ref{linea} we can also establish that $\tau_n \simeq 7 \tau_n$, which fits  well with the experimental
ratio 140/20.
An example of the dynamical characteristics can be better observed in
the numerical case of Fig. \ref{posneg}. Here, the time intensity profile of a single site of the
chain is compared with its driver neighbor in the pseudochaotic regime, for both
positive (Fig. \ref{posneg}(a)) and negative (Fig. \ref{posneg}(b)) coupling.
The chains have been chosen to have approximately the average period $<T>=0.95$ ms,
and $<T>=1.25$ respectively,  with $N=7$ in the negative case and $N=70$ in the positive.
\begin{figure*}
\begin{center}
\includegraphics[width=4cm,height=9cm]{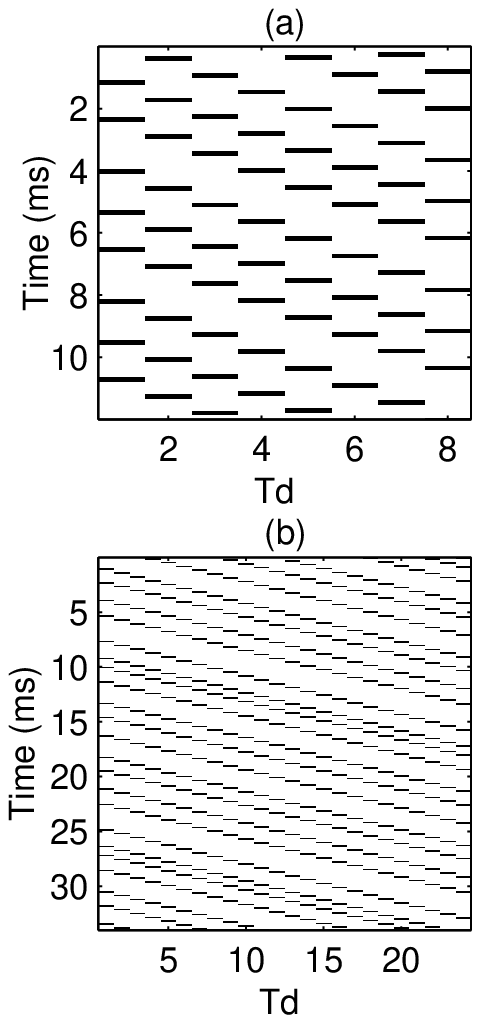}
\includegraphics[width=4cm,height=9cm]{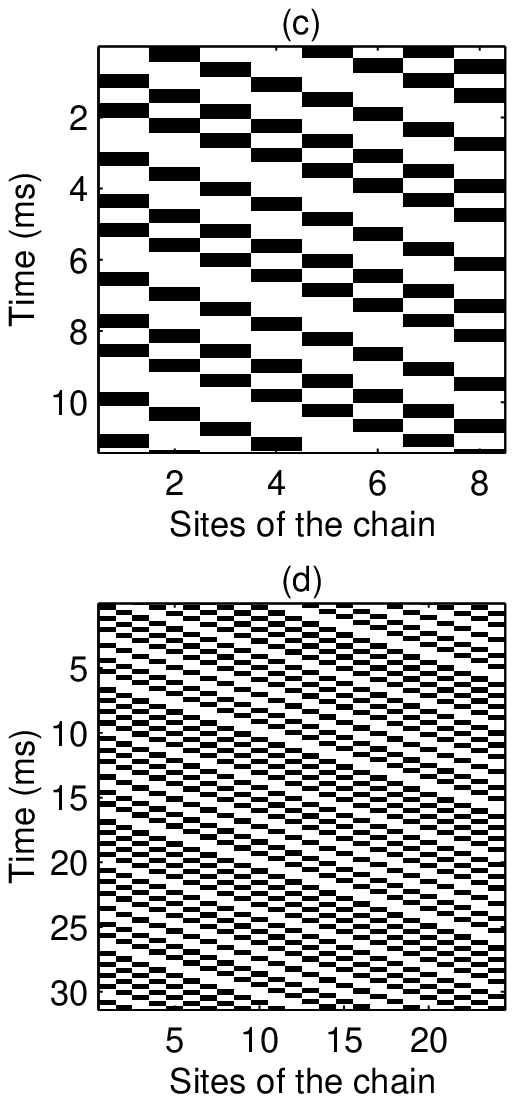}
\caption{Space time representation of the experimental laser intensity in the experiments of Ref. \cite{dss} (a,c) compared
with the numeric result of a oscillators chain (b,d). (a) T$_d$=4 ms, coupling strength -7\% (c) T$_d$=12 ms, coupling
strength -7\%. For the equivalent chain of oscillators: (b) N=8, $\epsilon=-0.042$ (d) N=24, $\epsilon=-0.042$}
\label{compare}
\end{center}
\end{figure*}

 We can already establish a full equivalence between the experimental laser system of Ref. \cite{dss} and the closed chain.
In Fig. \ref{compare} a chain of $N=8$ coupled negatively elements is compared to an experiment performed with negative feedback
and a delay time of $4$ $ms$, and another chain of $N=24$ elements with an experiment in which a delay of $12$ $ms$ was used.
 For the experimental case we show the Space Time Representation of the laser intensity \cite{STR} (Fig.\ref{compare}(a) and (c)),
and for the numerical case we report in gray scale the intensities for the oscillators as a function of time (Fig. \ref{compare} (b) and (d)) .
The experimental results correspond to a negative coupling of $-7\% $, while the numerical results are obtained for $\epsilon=-0.046$,
these values being close to the threshold values to establish a stable pattern in the dynamics.

 This same behavior in which a positive coupling can induce a collective synchronous behavior,
 while the negative coupling induces an antiphase dynamics has been observed in experiments
in arrays of neurons where the coupling could be modified  \cite{neuro_sig}.

In this work we demonstrated that in certain conditions a time delayed system can be used for an experimental
verification of the synchronization behavior of arrays of chaotic oscillators. In an example of this equivalence
for a delayed laser in the homoclinic chaotic regime, phase and antiphase pseudochaotics synchronization regimes have been reported.

%\end{multicols}
\end{document}